# Talbot effect in nonparaxial self-accelerating beams with electromagnetically induced transparency

JINGMIN RU,[1] ZHENKUN WU,[1, 2*] YAGANG ZHANG,[1]  FENG WEN,[3#] AND YUZONG GU[1]

[1]*Institute of nano/photon materials and application, School of Physics and Electronics, Henan University, Kaifeng 475004, PR China*

[2]*National Demonstration Center for Experimental Physics and Electronics Education, School of Physics and Electronics, Henan University, Kaifeng 475004, PR China*

[3]*Key Laboratory for Physical Electronics and Devices of the Ministry of Education & School of Science & Shaanxi Key Lab of Information Photonic Technique & Institute of Wide Bandgap Semiconductors, Xi'an Jiaotong University, Xi'an 710049, PR China*

Corresponding author: *wuzk1121@163.com; [#]fengwen@mail.xjtu.edu.cn

**Abstract:** In this study, we report on the fractional Talbot effect of nonparaxial self-accelerating beams in a multilevel electromagnetically induced transparency (EIT) atomic configuration, which, to the best of our knowledge, is the first study on this subject. The Talbot effect originates from superposed eigenmodes of the Helmholtz equation and forms in the EIT window in the presence of both linear and cubic susceptibilities. The Talbot effect can be realized by appropriately selecting the coefficients of the beam components. Our results indicate that the larger the radial difference between beam components, the stronger the interference between them, the smaller the Talbot angle is. The results of this study can be useful when studying optical imaging, optical measurements, and optical computing.

**Keywords** Multilevel atomic configuration; Nonparaxial self-accelerating beam; Talbot effect; Electromagnetically induced transparency

**PACS numbers** 03.65.Ge, 03.65.Sq, 42.25.Gy

## 1. Introduction

Atomic systems are special non-solid materials that have been the subject of substantial research and development in recent decades. Studies have shown that such ideal media can exhibit controllable optical properties, which gives them a wide range of potential applications. Due to its strong dispersion and free-absorption properties, electromagnetically induced transparency (EIT) [1, 2] has been studied in atomic vapors and it was important in the development of multi-wave mixing processes [3-6]. For example, in 1995, Hemmer et al. observed enhanced four-wave mixing based on EIT and studied its propagation in a four-level double-Λ atomic system [7, 8]. In 2007, Zhang et al. demonstrated coexisting four- and six-wave mixing via two ladder-type EIT atomic vapors [9]. More recently, periodically-dressed atomic systems have been investigated and many interesting phenomena were observed. This included enhanced multi-wave mixing signals caused by Bragg reflection from photonic bandgap structures [10], edge solitons in photonic graphene [11], photonic topological insulators [12], (anti-) parity-time symmetric systems [13-15], optical Bloch oscillation, and Zener tunneling [16]. Moreover, atomic vapors are useful when studying paraxial nondiffracting Airy beams [17] and nonparaxial self-accelerating Bessel, Weber, and Mathieu beams [18] as its optical properties are well known.

In contrast, the Talbot effect was originally discovered by H. F. Talbot in 1836 [19], which displays a self-imaging of periodic objects without utilizing any optical components. This phenomenon is explained as a consequence of Fresnel diffraction and diffracted beams interfering by Lord Rayleigh in 1881 [20]. The peculiar properties of the self-imaging effect can be used for many applications such as optical testing, optical metrology, spectrometry, and synthesis. The Talbot effect has been extensively explored in fields such as atomic media [21-23], waveguide arrays [24], photonic lattices [25], Bose-Einstein condensates [26], and both paraxial and nonparaxial accelerating beams [27-29]. The Talbot effect can be achieved using second-harmonic generation [30], exciton polaritons [31], and rogue waves [32, 33], but in general it requires transverse periodicity of the input wave. In addition, study of Talbot effect is still an active topic, and recently the efficient trajectory manipulation of the Airy-Talbot effect in dynamic linear potentials, and the self-imaging effect for a superposition of the fundamental circular Airy beams have been reported in Refs [34, 35].

In this letter, we theoretically and numerically investigated the fractional Talbot effect of nonparaxial self-accelerating beams in a three-level nonlinear atomic medium exhibiting EIT, which, to the best of our knowledge, has not been undertaken in previous works. The fractional Talbot effect realized in multi-level atomic media can be tuned by modifying multiple parameters, including density of the atomic medium, and amplitude, phase, and frequency detuning of the coupling fields; therefore, the Talbot effect could be used in various applications in the fields of optical imaging, image processing, and synthesis.

## 2. Propagation model

In our proposed approach, first, a three-level atomic configuration of sodium is formed with an excited level state, $|2\rangle$ (4D$_{3/2}$), and two lower level states, $|1\rangle$ (3P$_{1/2}$) and $|0\rangle$ (3S$_{1/2}$), as demonstrated in the inset panel in figure 1(a). The pump field $\Omega_1$ and strong control filed $\Omega_2$ are used to couple the atomic transitions $|0\rangle \rightarrow |1\rangle$ and $|1\rangle \rightarrow |2\rangle$, respectively. We restrict our attention to the evolution of a continuous wave optical beam $\psi(r)$ propagating in the x-z plane, i.e., $\psi(r) = \psi(x,z,t)\hat{y}$. Derived from Maxwell's equations, the nonparaxial propagation of envelope $\psi$ inside the atomic vapor cell exhibiting Kerr nonlinearity is described by [18, 27, 36]

$$\left(\frac{\partial^2}{\partial z^2} + \frac{\partial^2}{\partial x^2}\right)\psi(x,z) + K^2 \psi(x,z) = 0. \tag{1}$$

Here, $x$ and $z$ are the transverse and longitudinal coordinates, $r = \sqrt{x^2 + z^2}$ and $K = k\sqrt{1+\chi}$. Here, $k$ is the wave number, while $\chi$ is the susceptibility of the atomic vapor system, which will compete with different order for various field intensities. We define susceptibility as $\chi = \chi_1 + \chi_3 |\Omega_2|^2$, where the linear $\chi_1$ and cubic $\chi_3$ terms, respectively, are expressed as [37-39]

$$\chi_1 = \frac{\eta}{\kappa} \tag{2}$$

and

$$\chi_3 = -\frac{\eta}{\kappa^2 d}. \tag{3}$$

Here, $\eta = iN\wp_{10}^2/\hbar\varepsilon_0$, $\kappa = (\Gamma_{10} + i\Delta_{10}) + |\Omega_2|^2/d$, and $d = \Gamma_{20} + i(\Delta_{10} + \Delta_{20})$; $\Omega_2 = \wp_{12}E_2/\hbar$ is the Rabi frequency of the control field; $\wp_{ij}$ is the transition electric dipole moment between levels $|i\rangle$ and $|j\rangle$; N is the atomic density; and $\Gamma_{ij}$ denotes the rate at which the population decays from state $|i\rangle$ to state $|j\rangle$. Finally, $\Delta_{10} = \omega_1 - \omega_{10}$ and $\Delta_{20} = \omega_2 - \omega_{21}$ are the pumping and coupling field detunings, respectively, in which the atomic resonant frequency and laser frequency are represented by $\omega_i (i=1,2)$ and $\omega_{ij}(j=0,1)$. We set the parameters to be $N=10^{13}$ cm$^{-3}$, $\wp_{10}=3\times10^{-29}$ C m, $\lambda=532$ nm, $\Gamma_{20}=2\pi\times0.485$ MHz, $\Gamma_{10}=2\pi\times4.86$ MHz, and $\Delta_{20}=0$ in our numerics for convenience without special statement [37-41].

## 3. Calculations and discussions

Figures 1(a) and 1(b) show the imaginary (orange curve) and real (blue curve) parts of $\chi$ against the pumping detuning for various controlling intensities; these correspond to the absence ($\Omega_2=0$) and presence ($\Omega_2=3.5\times10^7$ s$^{-1}$) of the control field, respectively. Figure 1(a) shows that when the control field $\Omega_2$ is absent, the atomic configuration has high absorption (the orange curve). However, when $\Omega_2$ is applied, an EIT window is generated under the two-photon resonance condition with $\Delta_{10}+\Delta_{20}=0$, as shown in figure 1(b). The absorption and dispersion properties of the medium are significantly different due to free-absorption and enhanced Kerr-nonlinear refraction (blue curve) near the two-photon atomic resonance $\Delta_{10}=\Delta_{20}=0$. These factors have a major effect on the formation and evolution of the self-imaging process.

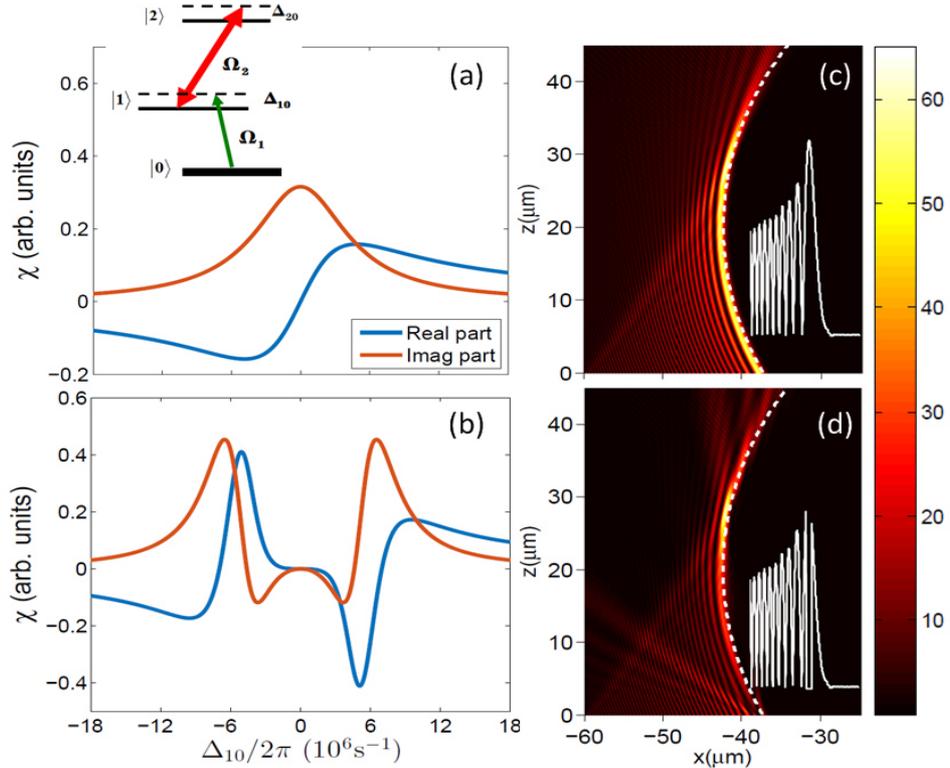

**Figure 1**. (Color online) The susceptibility versus the pumping frequency detuning at controlling

intensity (a) $\Omega_2 = 0$ and (b) $\Omega_2 = 3.5\times 10^7\ s^{-1}$; the inset in (a) shows the relevant ladder-type atomic system. (c) Nonparaxial accelerating beam in evolution with the theoretical accelerating trajectory indicated by the white dashed curve. (d) The same setup as (c), but the main lobe initially cut out. The color bar for the transverse plane is the same in all other figures.

A well-known exact solution of equation (1) in polar coordinates uses loss-proof self-accelerating beams [18, 42], which are eigenmodes of the Helmholtz equation in the atomic configuration. This has the general form

$$\psi(x,z) = \int_{-\pi/2}^{\pi/2} d\theta \{\exp(i\alpha\theta)\times\exp[iK(x\sin\theta + z\cos\theta)]\} \quad (4)$$

where $\alpha$ is a real parameter that determines the radius $\sim \alpha/K$ of the main lobe of the solution via $r \approx \alpha/K$. We consider a loss-proof self-accelerating beam generated at the two photon resonance point $\Delta_{10} = \Delta_{20} = 0$, which has a corresponding effective wavenumber of $K \approx 1.181\times 10^7 + i1.0742\ \mathrm{m}^{-1}$. Using the input described by equation (4) with $z = -20\ \mu\mathrm{m}$ and $\alpha = 500$, the corresponding evolution of nonparaxial self-accelerating beams can be displayed numerically using the Helmholtz equation, equation (1), directly under the EIT condition. The intensity profile of the beam against the propagation distance is presented in figure 1(c), and the inset represents the energy distribution of the beam at $z = 0$. The robust nonparaxial beam can propagate in the EIT window with much less absorption and the main lobe follows a circular trajectory, as indicated by the white dashed curve. In figure 1(d), the evolution of beam indicates self-healing properties. The main lobe of the beam is initially screened out, but it recovers quickly as the propagation distance increases due to energy transfer from the tail to the head of the beam.

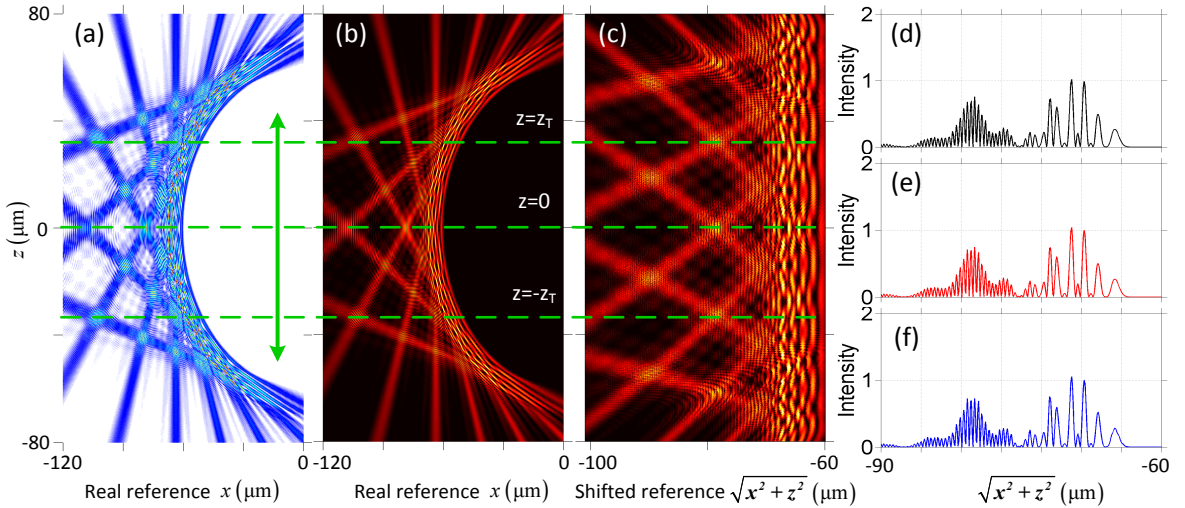

**Figure 2**. (Color online) Nonparaxial accelerating Talbot effect caused by the superposition of nonparaxial accelerating beams in (x, z) coordinates by analytical solutions (a) and numerical simulation (b). (c) Accelerating ($\sqrt{x^2+z^2}$, z) coordinates. Parameters: $\Delta\alpha = 20$, $c_\alpha = 1$, and $\alpha \in [700, 800]$. (d)-(f) Intensity cross section at three planes of self-imaging, $z = -z_T$, 0, and $z_T$, marked by dashed lines in (c). In order to make the comparison clearer, the cross sections are shifted along the circular trajectory.

Similarly to previous work, the propagation properties of the general solution can be verified as a superposition. Here, an input can be constructed by superposition of the accelerating solutions, as described by equation (4), with different order $\alpha$. That is

$$\psi(x,z) = \int_{-\pi/2}^{\pi/2} d\theta \left\{ \exp[iK(x\sin\theta + z\cos\theta)] \times \sum_{\alpha \in \mathbb{Z}} c_\alpha \exp(i\alpha\theta) \right\} \quad (5)$$

where $c_\alpha$ is the amplitude of each component. We execute the propagation numerically with parameter $c_\alpha \equiv 1$ for $\alpha$ from 700 to 800 in an EIT atomic configuration. The input associated with equation (5) can be written as

$$\psi^0(x,0) = \int_{-\pi/2}^{\pi/2} d\theta \left[ \exp(iKx\sin\theta) \times \sum_{\alpha \in \mathbb{Z}} c_\alpha \exp(i\alpha\theta) \right]. \quad (6)$$

The final expression for the field at the output plane is

$$\psi(x,z) = F^{-1}\left\{ F\{\psi^0(x,0)\} \times \exp\left[\left(-\sqrt{\omega^2 - K^2}\right)z\right] \right\}, \quad (7)$$

where $F$ and $F^{-1}$ are the Fourier transform and inverse Fourier transform, respectively. The parameters $\omega$ can be derived from

$$F\left[\psi^0(x,0)\right] = \int_{-\infty}^{+\infty}\left[\psi^0(x,0) \times \exp(-i\omega x)dx\right]. \quad (8)$$

A given beam structure at any propagation distance $z$ can be obtained numerically. Our results are summarized in figure 2, which shows the intensity of the evolution carpet for the nonparaxial accelerating Talbot effect in the real (x, z) (figure 2(b)) and accelerating ($\sqrt{x^2+z^2}$, z) (figure 2(c)) frames of reference, although there are some distortions. To make a comparison with figure 2(b), the corresponding analytical result by performing equation (5) directly is also represented in figure 2(a), from which one can see that the analytical and numerical results agree with each other very well. These figures show that there is an extended region where the nonparaxial acceleration Talbot effect is clearly indicated by the intensity of the propagating input beam. This is substantially different from past works, where the nonparaxial accelerating Talbot effect originated from eigenmodes of the Helmholtz equation and propagated in a nonlinear atomic configuration. During evolution the superposed beams interfere with each other and exhibit accelerating self-imaging properties along the $z$ direction. The physical explanation for the nonparaxial accelerating Talbot effect is clear: the component lobes undergo superposition, interact as they propagate, and accelerate in unison; the accelerating Talbot effect forms due to interference and synchronized acceleration of the composed beams [43]. To visualize the self-imaging property, the intensities of the beams during propagation are presented at three planes: $z = -z_T, 0, z_T$ (where $z_T$ is the Talbot length), marked by dashed lines in figure 2(c). The three plots, presented in figures 2(d)–(f), are shifted along the circular trajectory to align them to simplify the comparison. The diffraction pattern repeats at the Talbot length planes.

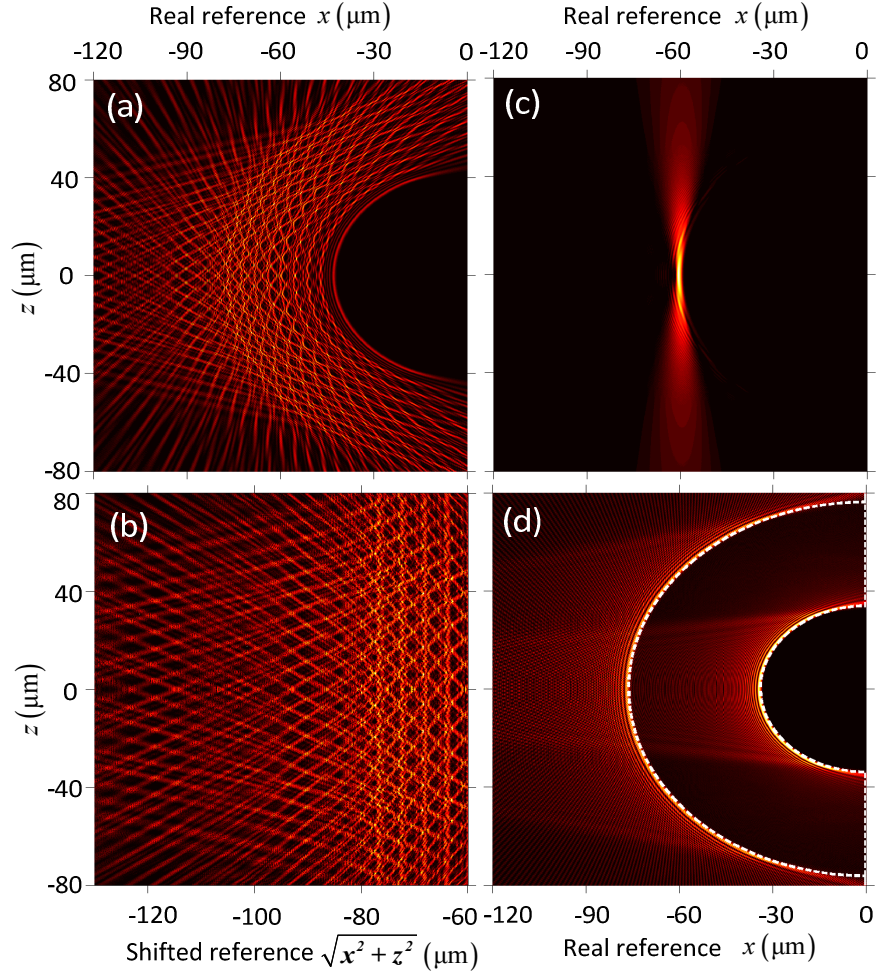

**Figure 3**. (Color online) Intensity distributions of the superposed nonparaxial accelerating beams.

(a) $500 \leq \alpha \leq 900$ with $\Delta\alpha=50$; (b) same as (a), but in the accelerating frame; (c) $700 \leq \alpha \leq 720$ with $\Delta\alpha=2$; (d) $m=400$ and $m=900$, the theoretical accelerating trajectory is indicated by the white dashed curve.

It is important to note that a distortion phenomenon is exhibited along with the Talbot effect during beam propagation. Phenomenologically, according to Huygens' principle, each point in the wavefront is a new source; thus, there will be an increasing number of sources and secondary beams as the beam evolves. These secondary beams are reflected, and thereby interfere with each other, resulting in the observed distortion.

The schematic diagrams in figure 2 show that the nonparaxial beams accelerate along circular trajectories in the EIT window. So, instead of the Talbot length, the Talbot angle can be used to explore the self-images. Mathematically, the two nearest components in equation (6) can be viewed as a combination of solutions with orders of $\alpha_0 + m\Delta\alpha$ and $\alpha_0 + (m+1)\Delta\alpha$. That is,

$$\exp[i(\alpha_0 + m\Delta\alpha)\theta] + \exp\{i[\alpha_0 + (m+1)\Delta\alpha]\theta\} \\ = \exp[i(\alpha_0 + m\Delta\alpha)\theta][1 + \exp(i\Delta\alpha\theta)], \qquad (9)$$

where $\alpha_0$, $m$, and $\Delta\alpha$ are the reference value of $\alpha$, an arbitrary integer, and the radial difference, respectively. If $\Delta\alpha\theta = 2n\pi$ where $n$ is a positive integer, the contribution from the nearest components will be the same because $\exp(i\Delta\alpha\theta) = 1$. Thus, equation (9) can be rewritten as $2\exp(i\alpha_0\theta)$, which is independent of $\alpha$. Therefore, an analytical expression of the Talbot angle is

$$\theta_T = \frac{2\pi}{\Delta\alpha}. \qquad (10)$$

From Equation (10), it is clear that the Talbot angle is independent of $\alpha$, and it decreases with increasing radial difference $\Delta\alpha$, i.e., the smaller $\Delta\alpha$ is, the larger the Talbot angle will be. Herein, we apply these results to some specific cases. Figures 3(a) and 3(b) show distributions composed of nine components for $500 \leq \alpha \leq 900$ with $\Delta\alpha = 50$ in the real and accelerating reference frames, respectively. It is noteworthy that the quality of resolution of the nonparaxial Talbot effect in Figures 3(a) and 3(b) is much improved. In particular, by increasing $\Delta\alpha$, a more precise accelerating Talbot carpet with a smaller Talbot angle can be obtained.

However, as shown in Figure 3(c), when the Talbot angle $\theta_T = \pi$, the superposition of nonparaxial accelerating beams does not yield the Talbot effect because beams propagate along a straight line without bending, which are increasingly similar to the interference of Gaussian-like beams. In Figure 3(d), we present the intensity distributions of two superposed nonparaxial accelerating beams for $400 \leq \alpha \leq 900$ with $\Delta\alpha = 500$; in this case, the component with $\alpha = 900$ has a considerably larger radius than the component with $\alpha = 400$. Thus, the interference between these two components is significantly weakened, and consequently, only the component with $\alpha = 400$ is retained as the component with continuously increasing intensity $\Delta\alpha = 500$. These results are qualitatively consistent with the observations in the previous studies [29, 36]. Based on the method discussed in [36], results reported in this Letter can easily be experimentally realized; e.g., a spatial light modulator (SLM) imposes the appropriate phase profile for generating nonparaxial accelerating beams, and two objective lenses (X60, NA=0.85) focus the accelerating beams to the micrometer scale.

Finally, we remark that the Talbot effect can be "constructed" by appropriately selecting coefficients $c_n$. As in previous works [27-29], we selected $c_n = [...,1,0,1,0,...]$ and $\Delta\alpha=10$ to reproduce the beam evolution in figure 2(b). The corresponding Talbot effect can also be formed so that the intensity carpet has approximately the same Talbot length in the real and accelerating frames, as depicted in figures 4(a)-4(c), respectively. Curiously, the intensity distribution Talbot angle shown in figure 4(b) is the same as that in figure 2(b), even though $\Delta\alpha=20$ in that case. Accordingly, because $c_n = [...,1,0,1,0,...]$, the coefficients of the odd components are 0 and the input superposition in equation (6) is $\exp(i\alpha_0\theta)\sum_{n\in\mathbb{Z}}\exp(i2n\Delta\alpha\theta)$. Hence the Talbot angle in this case reaches $\theta_T = \pi/\Delta\alpha$, so the problem the same as that presented in figure 2(b). Furthermore, at the half-Talbot length the beam recovers and there is a fractional nonparaxial accelerating Talbot effect in analogy to traditional Talbot effect [22, 28, 29].

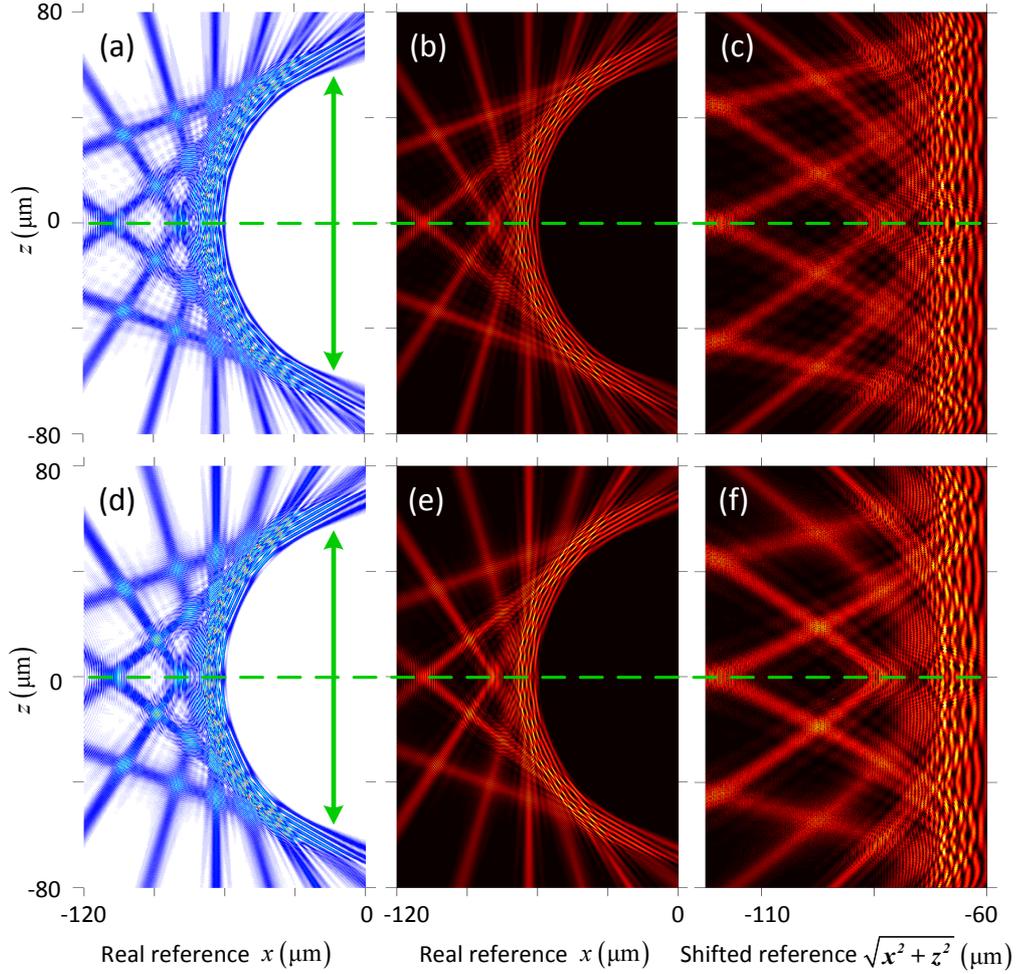

**Figure 4**. (Color online) Talbot carpets. Left column: intensity distributions of the superposed nonparaxial accelerating beams in $(x, z)$ frames by analytical solutions. Middle column: numerical simulation. Right column: numerical simulation in accelerating frames. Cases where (a)-(c) $c_n = [..., 1, 0, 1, 0, ...]$ and (d)-(f) $c_n = [..., 1, i, 1, i, ...]$. Parameters: $\Delta\alpha = 10$ for $\alpha \in [700, 800]$.

To clarify, the evolutions of input superposition $c_n = [..., 1, i, 1, i, ...]$ are presented in Figures 4(d)-4(f) wherein the setups are the same as those in Figures 4(a)-4(c), respectively. In particular, the propagations depicted in Figures 4(d)-4(f) are indications of the nonparaxial accelerating Talbot effect. Following the trend of the $c_n = [..., 1, 0, 1, 0, ...]$ case described above, here too, the Talbot angle is halved and can be given by $\theta_T = \pi/\Delta\alpha$. According to the analysis presented in [27-29], the Talbot effect for such special cases of coefficients is easy to understand.

In addition to EIT condition, we also examined the evolution under the conditions $\Delta_{10} = 4\,\text{MHz}$ and $\Delta_{20} = 0$ to indicate loss-proof property, as presented in figure 5. In this case, the system moved to the enhancement region [44, 45] yielding $K = 1.1807 \times 10^7 - i1.7916 \times 10^4 \ m^{-1}$. As shown in figure 5, the peak intensities of such self-accelerating beams in both cases are recorded upon propagation. It is clear that this condition corresponds to significantly larger absorption than under the two-photon resonance condition. In particular, we observed that the beam gradually attenuates as the propagation distance increases as long as the absorption in atomic vapor cell is large, while the beam maintains its peak intensity well at EIT condition. Consequently, one cannot obtain nonparaxial accelerating Talbot effect in such a case because the large absorption would have to be compensated by a correspondingly large energy transfer from the tail to the head of the beam, which would break the profile of an accelerating beam in evolution.

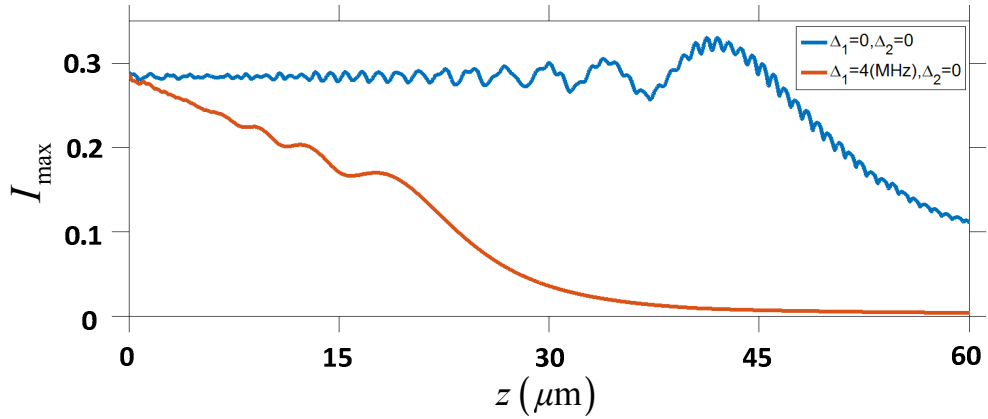

**Figure 5**. (Color online) Loss-proof property. Comparison between the peak intensity of the self-accelerating beam at EIT condition (blue curve) and beyond EIT condition (orange curve) during propagation.

## 4. Conclusion

In summary, we have introduced the Talbot effect for nonparaxial self-accelerating beams under an EIT condition in a three-level atomic vapor cell. The EIT-assisted fractional Talbot effect was demonstrated by propagating superposed eigenmodes of the Helmholtz equation in a nonlinear atomic configuration. This was based on the interference of a superposition of coherent nonparaxial accelerating beams and was greatly affected by EIT. The Talbot effect emerges when appropriate beam components are chosen. We also discovered that the Talbot angle is determined by the radial difference $\Delta \alpha$. When $\Delta \alpha$ is smaller there is greater interference between the beams, which results in a larger Talbot angle. These findings enrich our understanding of the Talbot effect family. Our findings introduce new properties of the Talbot effect, which provide new avenues for research and development of new methods for studying nonparaxial accelerating Talbot effect in atomic media, which may broaden its applications in various optical fields.

## Acknowledgements

National Natural Science Foundation of China (NNSFC) (61805068, 11747046, 61875053); China Postdoctoral Science Foundation (CPSF) (2017M620300); Natural Science Fund of Shaanxi Province (2018JQ6002); Science and Technology Department of Henan Province (202102210111).